\documentstyle[aps,pre,amsfonts,twocolumn]{revtex}

\renewcommand{\vec}[1]{{\mathbf #1}}

\input{psfig}

\begin{document}

\draft

\title{Diffusing-wave spectroscopy of nonergodic media$^1$}

\author{F. Scheffold}
\address{University of Fribourg, CH-1700 Fribourg, Switzerland}

\author{S. E. Skipetrov}
\address{Department of Physics, Moscow State University, 119899 Moscow, Russia}

\author{S. Romer}
\address{University of Fribourg, CH-1700 Fribourg, Switzerland}
\address{Polymer Institute, ETH Zurich, CH-8092 Zurich, Switzerland}

\author{P. Schurtenberger}
\address{University of Fribourg, CH-1700 Fribourg, Switzerland}

\date{\today}
\maketitle

\begin{abstract}
\begin{center}
{\bf Abstract} \end{center} We introduce an elegant method which
allows the application of diffusing-wave spectroscopy (DWS) to
nonergodic, solid-like samples. The method is based on the idea
that light transmitted through a sandwich of two turbid cells can
be considered ergodic even though only the second cell is ergodic.
If absorption and/or leakage of light take place at the interface
between the cells, we establish a so-called ``multiplication
rule'', which relates the intensity autocorrelation function of
light transmitted through the double-cell sandwich to the
autocorrelation functions of individual cells by a simple
multiplication. To test the proposed method, we perform a series of
DWS experiments using colloidal gels as model nonergodic media. Our
experimental data are consistent with the theoretical predictions,
allowing quantitative characterization of nonergodic media and
demonstrating the validity of the proposed technique.
\end{abstract}


\section{Introduction}
\label{intro} Diffusing-wave spectroscopy
(DWS)\footnotetext[1]{Accepted for publication in {\em Physical
Review E.}}
\cite{maret87,pine88,pine89,pine90a,weitz93a,weitz93b,kaplan94} is
an extension of ``conventional'' dynamic light scattering (DLS)
\cite{berne76,brown93} to the multiple scattering regime. The basic
idea of DWS is to use the autocorrelation function $g_2(\tau) =
\left< I(t) I(t+\tau) \right>_T/\left< I(t) \right>_T^2$ of the
light intensity $I(t)$ scattered by a turbid medium to study the
dynamics of scatterers in the medium.  Here $\left< \cdots
\right>_T$ denotes {\em time\/} averaging which can differ from the
{\em ensemble\/} averaging $\left< \cdots \right>_E$. Introduced in
1987 \cite{maret87}, the technique of DWS has rapidly evolved in
recent years, and is currently applied to study various types of
turbid media, such as colloidal suspensions
\cite{maret87,pine88,pine89,pine90a,weitz93a,weitz93b,maret97},
particle gels and ceramic green bodies
\cite{romer00,yodh93,urbanb00,urban00,wyss01}, emulsions
\cite{gang95,hebraud97,lisy99}, foams
\cite{durian91,earnshaw94,gopal95,hohler97,gopal97}, granular
\cite{menon97,durian00}, and biological
\cite{boas97a,palmer99,lohwasser99} media. It has been demonstrated
that DWS can be used to image macroscopic static and dynamic
heterogeneities in turbid media
\cite{boas95,heck96,skip96,heck97,boas97,skip98,heck98}. In 1995,
Mason and Weitz have suggested that the motion of colloidal
particles, characterized by DWS, can be directly related to the
viscoelasticity of the surrounding medium (the corresponding
experimental technique is sometimes called ``DWS-microrheology'')
\cite{mason95}. Much attention has been paid to this approach since
it could provide fast and non-invasive access to viscoelastic
properties of numerous materials, thereby opening a large new field
of potential applications for DWS
\cite{urban00,gisler98,crocker00,mason00}.

One of the main reasons for the remarkable success of DWS is the
availability of a relatively simple and reliable theoretical model
which describes the experimental data in all practically important
cases.  The theoretical model of DWS is based on the diffusion
approximation \cite{pine89,pine90a}. It applies equally well to
both statistically homogeneous and heterogeneous media (media with
inclusions, scatterer flows, etc.).  Absorption of light,
reflection of scattered waves at the sample boundaries, modulation
of the source intensity, as well as various types of scatterer
motion (Brownian\cite{pine89,pine90a} and
sub-brownian\cite{weitz89} motion, laminar
\cite{pine90b,bicout91,bicout93} and turbulent\cite{bicout95}
flows, etc.)  can be taken into account within the framework of the
diffusion model. Recently, the theory of DWS has been extended to
nonlinear random media \cite{skip00,skip01}.

An important condition for the applicability of the existing
diffusion theory to DWS experiments is the {\em ergodicity\/} of
the turbid medium under investigation. Indeed, ensemble-averaged
quantities are commonly calculated theoretically, while it is the
time averaging which is most easily obtained in experiments. Thus,
$\left< \cdots \right>_E = \left< \cdots \right>_T$ is required for
the experimental data to be described by the theory.  If the
light-scattering sample is nonergodic (say, the sample or some part
of it is solid-like) additional efforts, e.g. translational or
rotational motion of the sample during the measurement, are
necessary in order to obtain $\left< \cdots
\right>_E$\cite{boas95,boas97,skip98,heck98}. Similar arguments
hold for the role of nonergodicity in standard DLS experiments
\cite{pusey89,joosten90,megen91,xue92,schatzel93,kirsch96,bartsch97}.

In the present paper we propose a novel, elegant and simple way of
performing diffusing-wave spectroscopy of nonergodic media
\cite{romer00,patent2000}. The method is based on the idea that
light transmitted trough a sandwich of two turbid cells can be
considered ergodic even if only the second cell is ergodic. We show
that the resulting transmitted multiple-scattered intensity is
ergodic despite the nonergodicity of random medium in the first
cell. Consequently, the usual DWS theory applies to the description
of the intensity autocorrelation function $g_2(\tau)$. Moreover,
the double-cell sample can be constructed in a way that the
presence of the second, ergodic cell does not obscure the
light-scattering signal of the nonergodic medium under
investigation.  In order to simplify the interpretation of the
experimental data, we introduce a so-called ``multiplication
rule''.  Namely, we show that if a significant attenuation of
diffuse waves occurs at the interface between the cells due to
absorption and/or leakage of light in transverse directions, the
ensemble-averaged {\em field\/}\cite{field} autocorrelation
function $g_1^{(2)} (L_1 ,L_2 ,\tau )$ of the double-cell setup
equals a product of autocorrelation functions $g_1^{(1)} (L_1 ,\tau
) $ and $g_1^{(1)} (L_2 ,\tau )$ corresponding to the individual
cells [from here on, we use superscripts $(1)$ and $(2)$ to denote
the autocorrelation functions corresponding to single- and
double-cell samples, respectively; $L_1$ and $L_2$ are the
thicknesses of the cells]. This allows a full DWS study of the
turbid media in the first cell despite its nonergodicity. Due to
its simplicity and high statistical accuracy, our method is ideally
suited for the study of turbid gels, colloidal glasses, and other
nonergodic multiple-scattering media
\cite{romer00,yodh93,wyss01,patent2000}. The method also extends
the applicability of the above-mentioned microrheology approach
\cite{mason95,gisler98} to an important class of solid-like complex
materials where the motion of colloidal tracer particles is highly
constrained.

The paper is organized as follows. In the next section, we provide
a brief review of known experiments in the field of DLS by
nonergodic media.  We describe the methods used to overcome the
problem of nonergodicity, and discuss the applicability of similar
methods in DWS experiments.  In the second part of the section, we
introduce an original, novel method to deal with nonergodicity in
DWS and reveal the conditions under which the method can be
applied.  Section \ref{theory} is devoted to the theoretical model
of DWS in a double-layer medium.  Starting from the diffusion
equation for the field autocorrelation function, we obtain an
expression for the autocorrelation function of light transmitted
through a sandwich of two turbid cells, separated by a
nonscattering but perhaps absorbing wall. Section \ref{mult}
introduces the so-called multiplication rule, and reveals the
conditions of its applicability.  In Sec.\ \ref{setup} we describe
our experimental setup, and Sec.\ \ref{results} is devoted to the
discussion of the main results of our experiments. First, we have
performed model experiments to test the validity of our theoretical
model, and to ensure that the experimental setup is adequately
described by the theory. In these experiments, no attenuation
occurs on passage of light through the wall separating the
light-scattering cells, and the multiplication rule does not apply.
Next, we show how our method can be applied to an important case of
considerable leakage of light at the inter-cell wall. This
situation is the most interesting for practical applications, since
the multiplication rule implies that the autocorrelation function
$g_1^{(1)}(L_1 ,\tau )$ corresponding to the first layer can be
obtained simply by dividing $g_1^{(2)}(L_1, L_2, \tau)$ measured
for the two-cell setup by $g_1^{(1)}(L_2 ,\tau)$ measured for the
second cell taken alone. In addition, our second sample is highly
asymmetric: the first layer is of significant optical thickness and
has a relatively short correlation decay time $\tau_1$, while the
optical thickness of the second one is moderate and its correlation
decay time $\tau_2 \gg \tau_1$. This prevents the light-scattering
signal of the first, nonergodic cell from being obscured by the
second cell. In Sec.\ \ref{apply} we discuss different aspects of
application and optimization of the two-cell technique in practice.
We summarize our main results in Sec.\ \ref{concl}.

\section{Dynamic light scattering by nonergodic media}
\label{ergodicity} 
\begin{figure}[t]
\vspace{0.5cm} \psfig{file=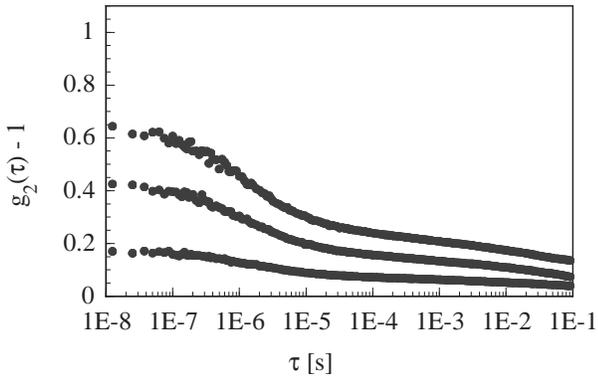,width=8cm} \vspace{0.5cm}
\caption{Typical light scattering signal of a non-ergodic system.
Repeated (time averaged) measurements of $g_{2}(\tau )-1$ lead to
non-reproducible results, each being of limited use for the
characterization of the system. Over the measurement time
(typically a few minutes), the time and ensemble averages of the
fluctuations of the scattered light intensity are not the same.}
\label{nonergod}
\end{figure}

In nonergodic random media, the scatterers are
localized near fixed average positions, probing only a small
fraction of their possible spatial configurations by thermal
motion. As a consequence, the measured time-averaged quantities
(such as the scattered intensity or its autocorrelation function)
differ from the ensemble-averaged ones. Experimentally, one finds
that a series of (time-averaged) measurements on a given sample
yields a set of different results, each being of limited use for
the characterization of the medium (see Fig.\ \ref{nonergod}).

\subsection{Concepts to deal with nonergodicity in
dynamic light scattering} \label{concepts} For diluted, nonergodic
samples several methods have been applied to properly average the
data obtained in DLS experiments
\cite{pusey89,joosten90,megen91,xue92,schatzel93,kirsch96,bartsch97}.
A good comparative review of some of these methods can be found in
Refs.\ \onlinecite{xue92} and \onlinecite{schatzel93}. The most
direct method of performing ensemble averaging of scattered light
for nonergodic samples is based on the idea of collecting light
scattered by different parts of the sample, thus performing the
``real'' ensemble averaging \cite{xue92}. Experimentally, the
sample is slowly moved or rotated, while the autocorrelation
function of the scattered intensity $g_2(\tau)$ is collected.

\noindent Obviously, this leads to an additional decay of
$g_2(\tau)$, which becomes increasingly rapid with increase of the
translation velocity or the rotation frequency \cite{xue92}. The
rotation/translation method can be extended to concentrated, turbid
suspensions without any particular difficulties, and it has been
actually employed in recent experiments on diffusing-wave imaging
\cite{skip98,heck98}. It has, however, an important disadvantage of
experimental complexity (translation or rotation of the sample is
required). Besides, it is not well suited for fragile turbid
systems (such as gels), since moving or rotating the sample can
lead to its disturbance or even breakdown.

Pusey and Van Megen \cite{pusey89} have proposed a method of
obtaining the ensemble-averaged intensity autocorrelation function
from a combination of static and dynamic experiments. The idea is
to measure the autocorrelation function of scattered intensity for
a single sample orientation, and to subsequently rotate/translate
the sample rapidly in order to obtain the average intensity for a
given scattering angle\cite{pusey89,joosten90,megen91,gang99}.
Recently, an extension of the above method to DWS experiments has
been developed\cite{nisato00}.

In general, all these approaches rely on a one dimensional motion
of the sample, i.e. scanning the speckle pattern, which is not very
efficient and therefore time consuming. This makes it intrinsically
difficult to apply these methods to systems which evolve in time,
e.g. gelling systems.

A different method of dealing with nonergodicity in DLS has been proposed in Refs. \onlinecite{kirsch96} and
\onlinecite{bartsch97} (see also Ref.\onlinecite{luca2000}). The authors use a CCD camera to record the temporal
evolution of many speckle spots simultaneously, which allows them to perform correctly both time and ensemble
averages (the latter is the average over a large number of speckle spots). The method appears to be very efficient
in the small-angle single-scattering regime, while its application to multiple-scattering systems is complicated
by a (generally) insufficient time resolution of available CCD cameras (in DWS, fast and low-intensity speckle
fluctuations are usually monitored). However, the method of Refs.\ \onlinecite{kirsch96} and
\onlinecite{bartsch97} can be applied to study extremely slow dynamics in turbid systems where the above mentioned
constraints do not apply\cite{harden2000}.

\subsection{Two-cell technique} \label{double}
\begin{figure}[t]
\vspace{0.5cm} \psfig{file=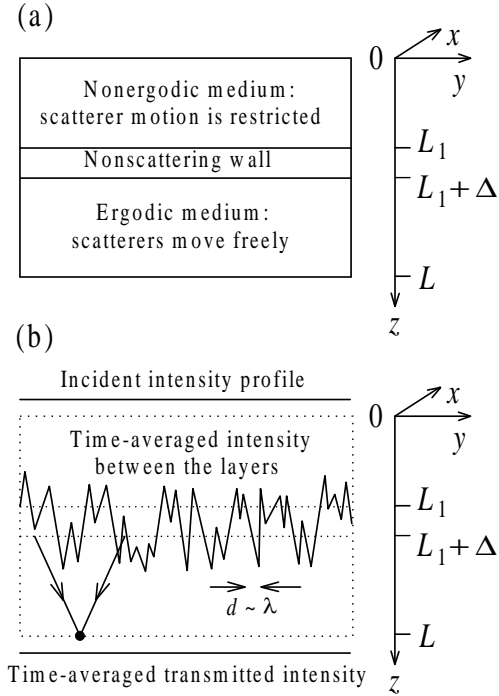,width=8cm} \vspace{-0.5cm}
\caption{Schematic illustration of the idea of the two-cell
technique. (a) A cell (thickness $L_2$) containing an ergodic
turbid medium is added just after the cell containing the
nonergodic medium under investigation (thickness $L_1$). The cells
are separated by a glass wall (thickness $\Delta$). The overall
thickness of the sample is $L = L_1 + \Delta + L_2$. (b)
Time-averaged intensity profiles are shown schematically for the
incident light, the multiple-scattered light at the inter-cell
interface $z = L_1$, and for the transmitted light. Between the
cells, the time-averaged intensity exhibits speckles due to the
nonergodic nature of the medium in the first cell. The speckles are
washed out upon the transmission through a sufficiently thick
second cell.} \label{sketch}
\end{figure}

To overcome the problem of nonergodicity, we prepare a sandwich
consisting of two independent glass cells. The first cell contains
a solid-like nonergodic medium under study, while the second cell
is filled with an ergodic medium [see Fig.\ \ref{sketch}(a)]. By
adjusting the concentration of scatterers in the second cell, the
viscosity of the liquid where the scatterers are suspended, and the
thickness of the cell, it is possible to shift the ``forced'' decay
of the autocorrelation function due to the second cell to long
correlation times $\tau$. Then $g_1^{(2)} (L_1 ,L_2 ,\tau ) $ will
exhibit a short-time decay due to the motion of scatterers in the
first cell, at intermediate $\tau$ it will saturate at a plateau
because of the nonergodic nature of the medium in the first cell,
and finally, at long correlation times, $g_1^{(2)} (L_1 ,L_2 ,\tau
) $ will decrease to zero due to the motion of scatterers in the
second cell.  Physically, the slow motion of scatterers in the
second cell gently shakes and randomizes the speckle pattern of the
nonergodic medium.  In this way, ensemble averaging of light
scattered by the first cell is accomplished, similarly to the case
when the sample is moved. The time scale of this averaging can be
well controlled by the scatterer motion inside the second cell, and
by the thickness of the cell.  The only constraint being a high
enough optical thickness of the second cell, ensuring that the
nonergodic intensity fluctuations produced by the first cell are
averaged out upon the transmission through the second one, and that
no light can pass through the latter without being scattered. The
described setup offers an advantage of high statistical accuracy
due to a two dimensional averaging scheme. Furthermore, no
mechanical disturbance is applied to the sample, which is of
particular importance when working with fragile systems.

Let us briefly discuss different aspects of time and ensemble
averages in the double-cell geometry. For simplicity, we choose the
intensity of light (and not its autocorrelation function) as an
object of averaging, but the same arguments apply to the
autocorrelation function as well. First, we consider the
transmission of a plane wave through a layer of {\em totally
rigid\/} random medium (thickness $L_1$, photon transport mean free
path $l_1^*$). We find that the ensemble-averaged intensity of
multiple-scattered light $\left< I(\vec{R}, z) \right>_E$ at depth
$z$ and transverse position $\vec{R} = \left\{ x, y \right\}$ is
independent of $\vec{R}$. Meanwhile, $\left< I(\vec{R}, z)
\right>_T \equiv I(\vec{R}, z)$ is a random function of $\vec{R}$,
a ``fingerprint'' of a given scatterer distribution in the sample.

\noindent Consequently, $\left< I(\vec{R}, z) \right>_E \neq \left<
I(\vec{R}, z) \right>_T$, and the considered light-scattering
system is nonergodic.

Let us now add a second turbid layer (thickness $L_2$, photon
transport mean free path $l_2^*$) just after the first one [we
neglect for a moment the thickness $\Delta$ of a glass wall
separating the cells, see Fig.\ \ref{sketch}(a)]. We assume that
the scatterers in the second layer are moving, and that all
possible spatial configurations of scatterers are explored in
course of this motion. The light intensity $I(\vec{R}, z, t)$ is
then time-dependent, and the time-averaged intensity of light
transmitted through the double-layer medium, $\left< I(\vec{R}, L,
t) \right>_T$, can be considered as a result of transmission of the
speckle pattern $\left< I(\vec{R}, L_1, t) \right>_T$, being
specific for the given configuration of scatterers inside the first
layer, through the second layer.  Now we assume that the optical
thicknesses of both layers are large enough ($L_n \gg l_n^*$), so
that the diffusion picture of wave transport is valid in both
layers.  This allows us to estimate the typical speckle size at the
$z = L_1$ plane as $d \sim \lambda$ (see, e.g., Ref.\
\onlinecite{shapiro86}), where the wavelength of light $\lambda$ is
assumed to be roughly the same in both layers and much smaller than
the scattering lengths in the layers. At the same time, as long as
$L_2 \gg l_2^*$, the {\em average\/} intensity of light transmitted
through the second layer is not sensitive to those details of the
``source distribution'' $I(\vec{R}, L_1, t)$ which are finer than
$L_2$. Consequently, the speckle pattern $I(\vec{R}, L_1, t)$ will
be completely washed out upon the transmission through the second
layer, and $\left< I(\vec{R}, L, t) \right>_T$ will be insensitive
to the exact spatial configuration of scatterers in the first
(rigid) layer, being equal to $\left< I(\vec{R}, L, t) \right>_E$.
The scattering system is then ergodic: the time and ensemble
averages are equivalent. Similar arguments apply in the case when
the first layer is not completely rigid, but exhibits some
nonergodic dynamics. The general reason for this is the following.
A given speckle spot of typical size $d \ll l_2^*$ in the $z = L_1$
plane can be considered as a point source of light, which produces
a photon cloud spreading linearly upon the diffusion through the
second layer of turbid medium. Hence, the speckle spot at $z = L_1$
gives contributions to the average intensity within a region of
typical transverse size $L_2$ at the $z = L$ plane.  For this
reason, the intensity of light at a given point $\vec{R}$ of the $z
= L$ plane is formed as a sum of contributions of $N \sim (L_2/d)^2
\gg 1$ speckle spots located at $z = L_1$ [see Fig.\
\ref{sketch}(b)]. In this way, the averaging over a large number of
independent speckle spots is performed. In some sense, such an
averaging is equivalent to that obtained by translation (or
rotation) of a single nonergodic layer. It is worthnoting, however,
that in order to average over the same number $N$ of speckle spots,
the sample has to be translated by a distance of order $L_2^2/d \gg
L_2$.

\section{Theory of DWS in a double-layer medium}
\label{theory} Adding a second cell modifies the intensity
autocorrelation function of transmitted light, which now exhibits
an additional decay. In order to interpret properly the
experimental data, we need a theoretical model describing the
autocorrelation function in a double-layer turbid medium. In this
section, we show that such a model can be readily constructed,
provided that the transport of light is diffusive in both layers.

Consider a slab situated between the planes $z=0$, $z = L = L_1 +
\Delta + L_2$, and consisting of two layers of turbid media
(thicknesses $L_1$ and $L_2$) separated by a nonscattering but
perhaps absorbing wall (thickness $\Delta$, absorption coefficient
$M_a$), as depicted in Fig.\ \ref{sketch}(a). $L_1$ and $L_2$, are
assumed to be much larger than the photon transport mean free paths
inside the layers, $l_1^*$ and $l_2^*$, respectively. If the slab
is illuminated by a plane monochromatic wave, the field correlation
function $G_1(\vec{r}, \tau)=\left< E(\vec{r}, t) E^*(\vec{r},
t+\tau) \right>_E$ inside each turbid layer can be found as a
solution of the diffusion equation \cite{boas95,skip96}:
\begin{eqnarray} \left[ \nabla^2 - \alpha^2(\tau) \right]
G_1(\vec{r}, \tau) = -\frac{3 S(\vec{r})}{l^*},
\label{difeq}
\end{eqnarray} where $\alpha^2(\tau)$ describes ``attenuation'' of correlation due to scatterer motion, and
$S(\vec{r}) = s_0 \delta(z-z_0)$ is the source term ($z_0 \approx
l_1^*$). The explicit form of the ``attenuation'' coefficient
$\alpha^2(\tau)$ in Eq.\ (\ref{difeq}) depends on the type of
scatterer motion (e.g., Brownian motion, flow, etc.). One can show
that ¨$\alpha^2(\tau)$ is determined by the absorption length $l_a$
and the mean square scatterer displacement $\left< \Delta r(\tau)^2
\right>_E$ \cite{weitz93a,skip96,correlation}:
\begin{eqnarray} \alpha^2(\tau) = \frac{3}{l_a l^*} +
\frac{k^2}{l^{*2}} \left< \Delta r(\tau)^2 \right>_E, \label{alpha}
\end{eqnarray} where $k=2\pi/\lambda$ is the wave number of light
in the medium.  For Brownian motion \cite{pine89,pine90a}, $\left<
\Delta r(\tau)^2 \right>_E = \left< \Delta r(\tau)^2 \right>_T = 6
D_B \tau$, and $\alpha^2(\tau) = 3/(l_a l^*)+6\tau/( \tau_0
l^{*2})$ with $\tau_0 = (k^2 D_B)^{-1}$, while for a directed flow
one finds\cite{pine90b,bicout91,bicout93} $\left< \Delta r(\tau)^2
\right>_{E ,T} \propto \tau^2$, and $\alpha^2(\tau) \propto
\tau^2$.  In the case of a gel, which is an example of nonergodic
medium considered in this paper, we assume that the particles
undergo a sort of arrested subdiffusive motion
\cite{romer00,nisato00,krall98}:
\begin{eqnarray} \left< \Delta r(\tau)^2 \right>_{E, T} =
\delta^2 \left\{ 1 - \exp\left[ -\left( \frac{\tau}{\tau_c}
\right)^p \right] \right\}, \label{arrested}
\end{eqnarray} where $\delta$ denotes the maximum r.m.s. particle displacement, $\tau_c$ is a characteristic
time required for this displacement to occur, and $p$ is a free
parameter.  We expect \cite{romer00} $p \simeq 0.7 \pm 0.1$. Eqs.\
(\ref{alpha}) and (\ref{arrested}) give
\begin{eqnarray}
\alpha^2(\tau) = \frac{3}{l_a l^*} + \frac{(k \delta)^2}{l^{*2}}
\left\{ 1 - \exp\left[ -\left( \frac{\tau}{\tau_c} \right)^p
\right] \right\}. \label{agel}
\end{eqnarray}

The time autocorrelation function for a sample consisting of two
turbid layers separated by a nonscattering wall can be found by
solving Eq.\ (\ref{difeq}) inside each layer, and then applying the
boundary conditions at the surfaces of the medium ($z = 0$ and $z =
L$) and at the interfaces between the turbid layers and the
nonscattering wall. Boundary conditions at $z = 0$ and $z = L$ are
of a well-known form \cite{zhu91,haskell94,aronson95,chen98}:
\begin{eqnarray}
&&\left[ G_1(\vec{r}, \tau) - z_1 \left(\vec{e}_z \cdot
\vec{\nabla} G_1(\vec{r}, \tau) \right) \right] \left|_{z =
0}\right.
= 0, \label{b1} \\
&&\left[ G_1(\vec{r}, \tau) + z_2 \left(\vec{e}_z \cdot
\vec{\nabla} G_1(\vec{r}, \tau) \right) \right] \left|_{z =
L}\right. = 0, \label{b2}
\end{eqnarray}
where $\vec{e}_z$ is a unit vector parallel to the $z$-axis, $z_n =
(2/3) l_n^*$, and the refractive index mismatch between the
scattering and transparent media is neglected. In general, a
rigorous theory would require using the extrapolation lengths $z_n$
calculated with account for the refractive index mismatch at the
sample surfaces \cite{zhu91,haskell94,aronson95,chen98}, or deduced
from the angular distribution of diffusely transmitted light
\cite{vera96}. For our purposes, however, it is sufficient to know
that $z_n \sim \ell_n^*$, since the actual values of $z_n$ are of
no importance in the limit of $\alpha_n l_n^* \ll 1$ and $l_n^*/L_n
\ll 1$ that we apply in the following.

Boundary conditions at the interfaces $z = L_1$ and $z = L_1 +
\Delta$ between turbid and nonscattering media are found by
applying the condition of flux conservation, as discussed by Ripoll
{\em et. al.\/} \cite{ripoll00}. Neglecting the refractive index
mismatch between the turbid and nonscattering media, we find
\cite{bound}:
\begin{eqnarray}
&&\left[ G_1(\vec{r}, \tau) + z_1 \left(\vec{e}_z \cdot
\vec{\nabla} G_1(\vec{r}, \tau) \right) \right] \left|_{z =
L_1}\right. \nonumber \\
&&\mbox{~~}= f(\vec{r}) \left[ G_1(\vec{r}, \tau) + z_2
\left(\vec{e}_z \cdot \vec{\nabla} G_1(\vec{r}, \tau) \right)
\right] \left|_{z = L_1 + \Delta}\right., \label{b3} \\
&&\left[ G_1(\vec{r}, \tau) - z_2 \left(\vec{e}_z \cdot
\vec{\nabla} G_1(\vec{r}, \tau) \right) \right] \left|_{z = L_1 +
\Delta }\right. \nonumber \\
&&\mbox{~~}= f(\vec{r}) \left[ G_1(\vec{r}, \tau) - z_1
\left(\vec{e}_z \cdot \vec{\nabla} G_1(\vec{r}, \tau) \right)
\right] \left|_{z = L_1}\right., \label{b4}
\end{eqnarray}
with $f(\vec{r})$ describing the losses of energy at the
inter-layer wall due to absorption and/or leakage of light in
transverse directions [$0 \leq f(\vec{r}) \leq 1$ and $f(\vec{r})
\equiv 1$ in the absence of both absorption and leakage]. For an
infinitely wide slab (no leakage) we have \cite{bound}
\begin{eqnarray}
f = \exp(-M_a \Delta) (1 - M_a \Delta) + (M_a \Delta)^2 \Gamma(0,
M_a \Delta), \label{fabs}
\end{eqnarray}
where $\Gamma(\ldots)$ is the incomplete gamma function. For
realistic samples of finite width, leakage of light in transverse
directions may be considerable. In the absence of absorption ($M_a
= 0$), we find for a cylindrical sample of radius $R \gg l_n^*$
after averaging over the sample crossection:
\begin{eqnarray}
f = 1 + \frac{1}{2}\left( \frac{\Delta}{R} \right)^2 - \frac{1}{2}
\frac{\Delta}{R} \left[4 + \left( \frac{\Delta}{R} \right)^2
\right]^{1/2}. \label{fleak}
\end{eqnarray}
Averaging over the sample crossection makes $f$ independent of
$\vec{r}$, which largely simplifies the further analysis without
affecting the final result qualitatively.

In what follows, we assume that $f(\vec{r})$ in Eqs.\ (\ref{b3}) and (\ref{b4}) is given either by Eq.\
(\ref{fabs}) or by Eq.\ (\ref{fleak}), being independent of $\vec{r}$ in both cases. The case when both absorption
and leakage are present can also be analyzed without any particular difficulties, but this leads to cumbersome
formulas without introducing any new qualitative features. Both Eqs.\ (\ref{fabs}) and (\ref{fleak}) exhibit a
monotonic decrease from $1$ ($M_a \Delta = 0$ or $\Delta/R = 0$) to $0$ ($M_a \Delta \gg 1$ or $\Delta/R \gg 1$).
If $f = 1$ (no absorption, no leakage), Eqs. (\ref{b3}) and (\ref{b4}) reduce to the well-known boundary conditions
between two turbid media in a direct contact \cite{boas95,skip96,heck97}. If $f = 0$ (strong absorption and/or
leakage), Eqs. (\ref{b3}) and (\ref{b4}) decouple and the problem is reduced to two single-layer problems.

Equations (\ref{difeq}) and (\ref{b1})--(\ref{b4}) enables us to
calculate the autocorrelation function of light transmitted through
the double-layer system, $G_1(L, \tau)/G_1(L, 0)$, which in the
absence of absorption ($l_a \rightarrow \infty$) inside the turbid
layers reads
\begin{eqnarray} &&g_1^{(2)}(L_1, L_2, \tau) =
\frac{\alpha_1 L_1}{\sinh(\alpha_1 L_1)} \frac{\alpha_2
L_2}{\sinh(\alpha_2 L_2)} \nonumber \\
&&\mbox{~~}\times \left( 1 - f^2 + 2 z_1/L_1 +
2 z_2/L_2 \right) \nonumber \\
&&\mbox{~~}\times \left[ 1 - f^2 + \frac{2 \alpha_1
z_1}{\tanh(\alpha_1 L_1)} + \frac{2 \alpha_2 z_2}{\tanh(\alpha_2
L_2)} \right]^{-1}, \label{g1two}
\end{eqnarray}
where we assumed $\alpha_n l_n^* \ll 1$ and $l_n^*/L_n \ll 1$.

In our experiments, the time-averaged autocorrelation function of
the scattered {\em intensity\/} $g_2(\vec{r}, \tau) = \left<
I(\vec{r}, t) I(\vec{r}, t+\tau) \right>_T/\left< I(\vec{r}, t)
\right>_T^2$ is measured.  As the double-layer system is shown to
be ergodic (see Sec.\ \ref{ergodicity}), $g_2$ is related to $g_1$
by the Siegert relation \cite{berne76}: \begin{eqnarray}
g_2(\vec{r}, \tau) = 1 + \beta \left| g_1(\vec{r}, \tau) \right|^2,
\label{siegert}
\end{eqnarray} where $\beta \simeq 1$ is a constant determined by
the experimental setup\cite{berne76,brown93}. Let us assume that
the first layer is filled with a gel, while the second one contains
a suspension of Brownian particles with a correlation decay time
$\tau_2 = \tau_0 (l_2^*/L_2)^2$. We also assume that the
characteristic correlation decay time of the first layer $\tau_1$
is much smaller than $\tau_2$. Then, it can be shown from Eqs.\
(\ref{g1two}) and (\ref{siegert}) that when the r.m.s. particle
displacement $\left< \Delta r(\tau)^2 \right>^{1/2}$ approaches
$\delta$ for $\tau \gg \tau_1$ [Eq.\ (\ref{arrested})], the
autocorrelation function $g_1^{(2)} (L_1 ,L_2 ,\tau )$ reaches a
plateau for $\tau_1 \ll \tau \ll \tau_2$. The plateau height can be
found from Eq.\ (\ref{g1two}). As $\tau$ approaches $\tau_2$,
$g_1^{(2)} (L_1 ,L_2 ,\tau )$ continues to decrease.

\section{Multiplication rule}
\label{mult} The autocorrelation function of light transmitted through a single layer of turbid medium can be
found\cite{skip96} from Eq.\ (\ref{difeq}) with boundary conditions (\ref{b1}) and (\ref{b2}), and since for
$\alpha l^* \ll 1$ and $L \gg l^*$ the solution reads
\begin{eqnarray}
g_1^{(1)} (L,\tau ) = \frac{{\alpha L}}{{\sinh (\alpha L)}}
\label{g1one}
\end{eqnarray}
We can rewrite Eq.\ (\ref{g1two}) as
\begin{eqnarray} g_1^{(2)}(L_1, L_2,\tau) &=& g_1^{(1)}(L_1, \tau)
\times g_1^{(1)}(L_2, \tau) \times F, \label{multi1}
\end{eqnarray}
where
\begin{eqnarray}
F &=&  \left( 1 - f^2 + 2 z_1/L_1 + 2 z_2/L_2 \right) \nonumber \\
&\times& \left[ 1 - f^2 + \frac{2 \alpha_1 z_1}{\tanh(\alpha_1
L_1)} + \frac{2 \alpha_2 z_2}{\tanh(\alpha_2 L_2)}
\right]^{-1}.\label{f}
\end{eqnarray}
Hence, the autocorrelation function of light transmitted through
the double-layer sample is given by a product of autocorrelation
functions of individual layers, times some function $F$ which
describes the coupling between the layers. By adjusting the
parameters of the experimental setup, $F \simeq 1$ can be achieved,
and then the following {\em multiplication rule\/} will hold:
\begin{eqnarray} g_1^{(2)}(L_1, L_2,\tau) &\simeq& g_1^{(1)}(L_1, \tau)
\times g_1^{(1)}(L_2, \tau). \label{multi2}
\end{eqnarray}
To reveal the conditions of validity of the multiplication rule
(\ref{multi2}), we consider separately the cases of low ($f \simeq
1$) and considerable ($f < 1$) losses of light at the inter-layer
interface.

If $f \simeq 1$, Eq.\ (\ref{f}) becomes
\begin{eqnarray}
F &\simeq&  \left[ \frac{L_1}{l_1^*} + \frac{L_2}{l_2^*} \right]
\nonumber \\
&\times& \left[ \frac{L_1}{l_1^*} \frac{\alpha_2
L_2}{\tanh(\alpha_2 L_2)} + \frac{L_2}{l_2^*} \frac{\alpha_1
L_1}{\tanh(\alpha_1 L_1)} \right]^{-1},\label{f1}
\end{eqnarray}
which reduces to unity only if $\alpha_1 L_1 \ll 1$ and $\alpha_2 L_2 \ll 1$, or $L_1/l_1^* \gg L_2/l_2^*$ and
$\alpha_2 L_2 \ll 1$, or $L_1/l_1^* \ll L_2/l_2^*$ and $\alpha_1 L_1 \ll 1$. In other words, for the
multiplication rule (\ref{multi2}) to hold the optical thickness of e.g. the first layer should be much greater
than that of the second one, while $g_1^{(1)}(L_2, \tau) \simeq 1$ for the latter.

In the presence of losses at the inter-layer interface, $f < 1$ and Eq.\ (\ref{f}) yields $F \simeq 1$ if
$L_n/l_n^* \gg 1/(1-f^2)$ and $\alpha_n l_n^*/\tanh(\alpha_n L_n) \ll 1 - f^2$. To give an example, suppose that
$50$\% of wave energy is lost on crossing the inter-layer interface once ($f = 0.5$). This gives $L_n/l_n^* \gg
4/3$ and $\alpha_n l_n^*/\tanh(\alpha_n L_n) \ll 3/4$. The latter conditions are commonly satisfied in typical DWS
experiments, making the application of the multiplication rule rather practical.

It is worthwhile to note that there exists a different, less
rigorous, but more transparent way of establishing the
multiplication rule.  Adopting the path-integral picture of light
propagation through the double-layer medium
\cite{pine88,pine89,pine90a}, we can write the field
autocorrelation function of transmitted light as
\begin{eqnarray} &&g_1^{(2)}(L_1, L_2, \tau) = \int_0^{\infty} d s_1 \int_0^{\infty}
d s_2\; P_2(s_1, s_2) \nonumber \\
&&\mbox{~~}\times \exp\left[ -\frac12 \left< \Delta\varphi^2(\tau)
\right>_{s_1} -\frac12 \left< \Delta\varphi^2(\tau) \right>_{s_2}
\right], \label{path2}
\end{eqnarray} where $P_2(s_1, s_2)$ is the relative weight of
paths consisting of segments of overall lengths $s_1$ and $s_2$
inside the first and the second layers, respectively.  The
variances of phase differences $\left< \Delta\varphi^2(\tau)
\right>_{s_n}$ describe the dephasing of light inside the layers.
For a single layer of thickness $L$ one
has\cite{pine88,pine89,pine90a}
\begin{eqnarray}
g_1^{(1)}(L, \tau) = \int_0^{\infty} d s\; P_1(s) \exp\left[
-\frac12 \left< \Delta\varphi^2(\tau) \right>_{s} \right].
\label{path1} \end{eqnarray} Obviously, if the variables $s_1$ and
$s_2$ can be considered independent, $P_2(s_1, s_2) = P_1(s_1)
P_1(s_2)$, and Eq.\ (\ref{path2}) reduces to a product of two
terms, each being of the form (\ref{path1}). This transparent
physical picture underlying the multiplication rule is particularly
helpful if one seeks to understand the role of losses (absorption
and/or leakage of light) at the inter-layer interface. Increasing
the losses makes the partial lengths of light paths $s_1$ and $s_2$
in the layers more and more independent, since the losses reduce
the probability for a typical path to cross the interface more than
once. Consequently, most of the paths will consist of two
independent segments, one inside the first layer (length $s_1$),
and the other one --- inside the second layer ($s_2$).  $P_2(s_1,
s_2)$ will then reduce to a product $P_1(s_1) P_1(s_2)$, exactly as
required for the multiplication rule to hold. Note that the Siegert
relation (\ref{siegert}) implies that the multiplication rule
applies to $g_2(\tau) - 1$ as well.

\section{Experimental setup}
\label{setup} 
\begin{figure}[t]
\vspace{0.5cm} \psfig{file=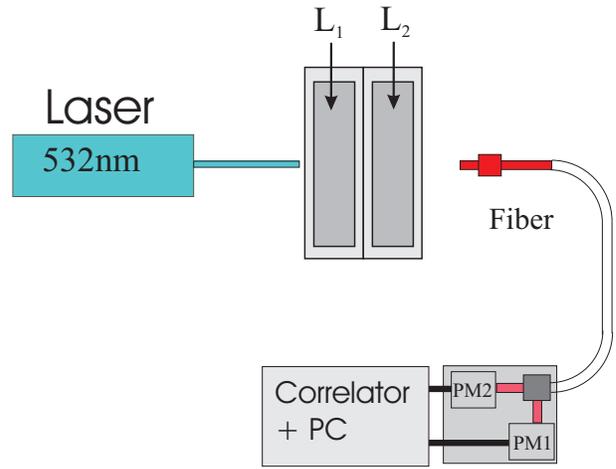,width=8cm} \vspace{0.5cm}
\caption{Experimental setup.  A laser beam ($\lambda_0 = 532$ nm)
is incident on a sandwich of two turbid layers. The light
transmitted through the sandwich is detected with a mono-mode
fiber, and is subsequently analyzed with a photomultiplier
(PM)/digital correlator unit.} \label{expsetup}
\end{figure}

Our experimental setup consists of a
frequency-doubled NdYAG laser (``Verdi'' from Coherent, wavelength
$\lambda_0 = 532$ nm) with a beam width of roughly $1$ mm. Only
minor differences are found upon expanding the beam to $7$ mm in
width (see also Ref.\ \onlinecite{kaplan94}). The laser radiation
is directed at the surface of the sample consisting of two cells,
and the multiple-scattered, transmitted light is collected using a
single mode optical fiber. The intensity of collected light is
analyzed by a digital correlator (see Fig.\ \ref{expsetup}). Apart
from the complex structure of the sample, the described
experimental setup represents a classical one for diffusing-wave
spectroscopy experiments in transmission geometry
\cite{pine88,pine89,pine90a,weitz93a}. We pay special attention to
the preparation of the sample, which is a key feature of our
experiments. The first cell (thickness $L_1$) is filled by a
colloidal gel, prepared from a destabilized solution of polystyrene
spheres (diameter $170$ nm) at a volume fraction up to $20$\% in a
buoyancy-matching mixture of water and heavy water
\cite{romer00,gauckler94}. In our experiments, the gel serves as a
model nonergodic medium where the average mean square displacement
of the constituent particles is reasonably well described by the
model of arrested subdiffusive motion [see Eq.\ (\ref{arrested})].
In general, gel systems undergo a complicated temporal evolution
(see, e.g., Refs.\
\onlinecite{romer00,urbanb00,urban00,wyss01,xue92,luca2000,krall98}),
which is not a subject of the present paper. For our measurements,
we wait after the sol-gel transition till the height of the plateau
of $g_2(\tau)-1$ is of the order $0.5$, and the gel properties
remain constant over the measurement time of about 10--30 min. We
note that experimentally $g_2(\tau)-1$ does not exhibit a perfect
plateau but rather a long stretching. This indicates a second,
long-time decay of the autocorrelation function, not included in
our simple theoretical model [Eq.\ (\ref{arrested})]. The
characteristic time scale of this decay extends to minutes or hours
\cite{romer00,luca2000}, and hence is beyond the time window of our
experiments.

Two different realizations of the two-cell sample were used in the
experiments reported below (samples A and B for brevity). While
sample A is a model sample constructed to test the validity of our
approach, sample B is optimized to facilitate its experimental
application (the multiplication rule holds for this sample).

Sample A. The sample consists of three disc-shaped microscope cover
plates (radius $R = 7$ mm, thickness $\Delta = 0.15$ mm) and two
ring-shaped teflon spacers.  The first cell contains a colloidal
gel (volume fraction $\Phi = 7$\%, photon transport mean free path
$l_1^* = 53$ $\mu$m). The second cell is filled with a suspension
of TiO$_2$ powder (Aldrich, particle diameter $< 5$ $\mu$m)
dispersed in pure glycerol (refractive index $n = 1.47$).

\noindent The transport mean free path for this suspension is
$l_2^* = 41$ $\mu$m.  The single scattering correlation time
corresponding to the suspension in the second cell is $\tau _0 = 23
\pm 1$ s, as determined by standard DWS measurements in
backscattering geometry\cite{pine89}.  The thicknesses $L_1$ and
$L_2$ of the layers were determined from DWS measurements in
transmission geometry by filling them with a suspension of known
$l^*$ and fitting the resulting intensity autocorrelation functions
with Eq.\ (\ref{g1one}). We find $L_1 = 1.75$ mm and $L_2 = 1.9$
mm. The optical thicknesses of both layers are comparable and high
enough ($L_1/l_1^* \simeq 33 \sim L_2/l_2^* \simeq 46 \gg 1$), and
therefore we call this sample ``symmetric''. A typical photon path
length $s$ in the sample is $s \sim L^2/l^* \approx 4 \times 10^2$
mm (with $L \simeq L_1 + L_2 \approx 4$ mm, $l^* = l_1^* \approx
l_2^* \approx 40$ $\mu$m), which is two orders of magnitude larger
than the sample thickness. Hence, the scattering of light in our
sample is essentially multiple, and a typical photon path is
expected to cross the buried inter-layer interface many times.
Estimation \cite{festimate} of the attenuation factor $f$ gives $f
\simeq 0.98 \simeq 1$, and hence the multiplication rule will only
hold for the sample A in a trivial situation of $\alpha_1 L_1 \ll
1$ and $\alpha_2 L_2 \ll 1$, as $L_1/l_1^* \sim L_2/l_2^*$.

Sample B. The sample consists of two equally thick light-scattering
cells (Hellma, $L_1 = L_2 = 1$ mm). The thickness of the glass wall
between the turbid media inside the cells is $\Delta \simeq 2$ mm.
Such a thick wall leads to a significant photon leakage in
transverse directions, and we put an aperture of radius $R = 2.5$
mm in between the two cells to prevent the photons leaked out of
the cell from reaching the detector. This leads to some additional
decrease of the total transmitted intensity, while ensures the
multiplication rule, as discussed in Sec.\ \ref{mult}. Indeed, for
$f \simeq 0.46$ estimated theoretically \cite{festimate} the
multiplication rule should hold whenever $L_n/l_n^* \gg 1/(1-f^2)
\sim 1$ and $\alpha_n \ell_n^* \ll 1-f^2 \sim 1$, which coincides
with the conditions of validity of Eq.\ (\ref{g1two}). We expect
the actual value of $f$ to be even smaller due to (total)
reflections of light at the interface which are not included in our
present theoretical model. Even though the conditions of validity
of the multiplication rule are satisfied, to apply it in a real
experiment [i.e. to find $g_2^{(1)}(L_1, \tau)-1$ as a ratio of
$g_2^{(2)}(L_1, L_2, \tau)-1$ and $g_2^{(1)}(L_2, \tau)-1$] we need
both $g_2^{(2)}(L_1, L_2, \tau)-1$ and $g_2^{(1)}(L_2, \tau)-1$ to
be essentially different from $0$, as otherwise experimental errors
in determination of the above autocorrelation functions can be
significant, making the application of the multiplication rule
impractical. To overcome this problem, we choose the second
light-scattering cell to be of much smaller optical thickness than
the first one, while the motion of scatterers is made slow inside
it. Namely, the first cell is filled with a very turbid gel (volume
fraction $\Phi = 20\%$, $l_1^* = 19$ $\mu$m), while the second one
contains a moderately turbid suspension of colloidal polystyrene
dispersed in a water-glycerol mixture ($l_2^* =$ 300--500 $\mu$m).
The decrease of the measured autocorrelation function due to the
motion of scatterers in the second, ergodic cell is thereby shifted
to long times $\tau \sim \tau_2 = \tau_0 (l_2^*/L_2)^2$ and does
not obscure the light-scattering signal of the first, nonergodic
cell. Hence, the multiplication rule can be efficiently applied for
$\tau \lesssim \tau_2$, while for $\tau \gg \tau_2$ the measured
autocorrelation function will decrease to $0$ due to the motion of
scatterers in the second cell. For the sample B, $L_1/l_1^* \approx
50 \gg L_2/l_2^* \approx$ 2--3, and therefore we call this sample
``asymmetric''.

\section{Results and discussion}
\label{results} In order to demonstrate the efficiency of the
method proposed and theoretically justified in Secs.\
\ref{ergodicity}--\ref{theory}, and to show the feasibility of
diffusing-wave spectroscopy in nonergodic media, we have carried
out several model experiments using samples A (symmetric) and B
(asymmetric).

\subsection{Symmetric sample}
\label{sym} 
\begin{figure}[t]
\vspace{0.5cm} \psfig{file=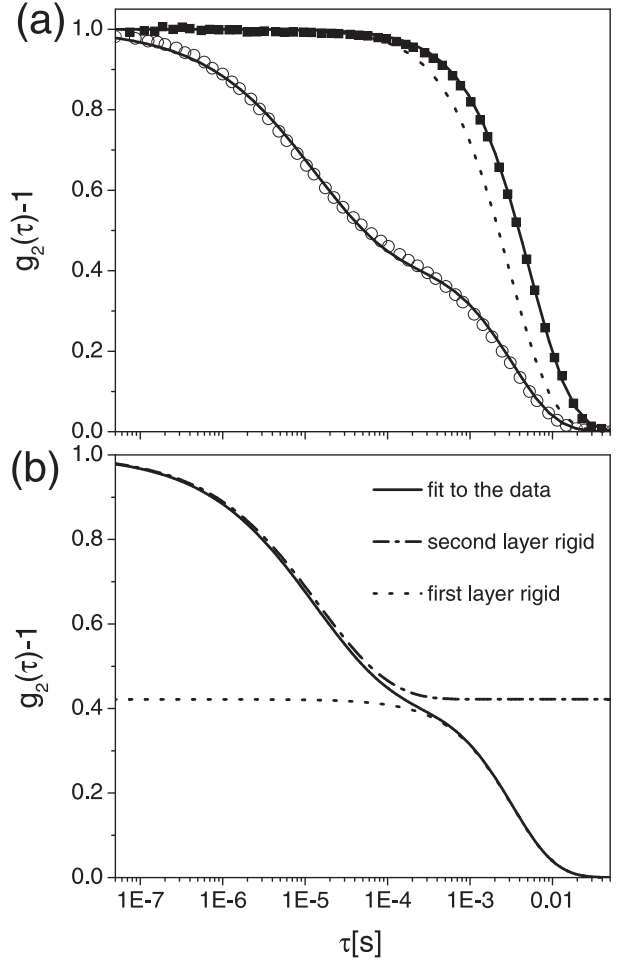,width=8cm} \vspace{0.5cm}
\caption{(a) Intensity autocorrelation function for transmission
through the symmetric two-cell sample A (open circles), and through
the isolated second cell (full squares). Theoretical fits [Eq.\
(\ref{g1two}) with $f = 1$ and Eq.\ (\ref{g1one})] are shown by
solid lines. The dotted line shows the theoretical correlation
function of the two-cell sample with the first layer assumed to be
rigid ($\tau_c \rightarrow \infty$). (b) Different contributions to
the intensity autocorrelation function are shown. The solid line is
the fit to the two-cell data of the panel (a). To obtain the
dashed-dotted line, we keep the second layer rigid ($\tau_0
\rightarrow \infty$), while all other parameters are unchanged. The
dotted line is that of the panel (a) multiplied by the plateau
value $0.422$.} \label{symmetric}
\end{figure}

Open circles in Fig.\ \ref{symmetric}(a) show the
intensity autocorrelation function $g_2^{(2)}(L_1, L_2, \tau) - 1$
of light transmitted through the symmetric sample A. The solid line
going through the open circles of Fig.\ \ref{symmetric}(a) is
obtained by fitting the experimental data with Eq.\ (\ref{g1two})
assuming $f = 1$ (no leakage of light at the wall separating the
cells). The fitting parameters are the gel parameters $\delta =
2.24$ nm, $\tau_{c}=25$ $\mu$s, $p=0.6$ [see Eq.\
(\ref{arrested})], while the parameters of the medium in the second
cell, as well as the cell thicknesses $L_1$ and $L_2$ were
determined independently. We note that the two sets of parameters
are only weakly correlated in the fit. Theory and experimental data
are found in excellent agreement.

As shown in Fig.\ \ref{symmetric}(a), the autocorrelation function
of the two-cell setup $g_2^{(2)} (L_1 ,L_2 ,\tau )-1$ exhibits two
characteristic decay times. The fast decay is due to the gel in the
first layer, while the slow decay corresponds to the dynamics in
the second layer. This is illustrated in Fig.\ \ref{symmetric}(b),
where we decompose $g_2^{(2)} (L_1 ,L_2 ,\tau )-1$ in two
contributions due to the dynamics in the first and second layers,
respectively. The contributions of the first (dash-dotted line) and
second (dotted line) layers are obtained by assuming the second or
the first layer to be rigid [Eq.\ (\ref{g1two}) with $f = 1$ and
$\alpha_2 \equiv 0$ or $\alpha_1 \equiv 0$, respectively]. The
contribution of the second layer is renormalized so that its value
at $\tau = 0$ is equal to the contribution of the first layer at
$\tau \rightarrow \infty$. The figure suggests that a simple
interpretation of the correlation function resulting from the
two-cell setup in terms of contributions of individual layers can
be given if the decay times due to different layers are well
separated.

The dotted line in Fig.\ \ref{symmetric}(a) shows the contribution
of the second layer to $g_2^{(2)} (L_1 ,L_2 ,\tau )-1$ [the same as
in Fig.\ \ref{symmetric}(b) but renormalized to 1]. Due to the
presence of the first layer, although assumed to be rigid, the
decay of the autocorrelation function becomes significantly faster
as compared to the second layer taken alone [full squares in Fig.\
\ref{symmetric}(a)]. This stems from the fact that the typical
lengths of photon paths in the second layer are increased due to
the presence of the first layer, which acts as an effective
``diffuse mirror'', increasing the probability for a light path to
get back to the second layer instead of being diffusely reflected
from the sample.

The measurements performed on the symmetric sample A allows us to
conclude that the multiple scattering of light in a two-cell sample
is correctly described by the theory developed in Sec.\
\ref{theory}. At the same time, the results of this subsection
serve as a test of our method, allowing diffusing-wave spectroscopy
to be applied to nonergodic random media. Indeed, the parameters of
the (nonergodic) gel ($\delta$, $\tau_c$ and $p$) can be obtained
from the fit to the experimental data [see Fig.\ \ref{symmetric}],
provided that the parameters of the medium in the second cell as
well as the thicknesses of both cells are measured independently.

\subsection{Asymmetric sample}
\label{asym} Although the results of the previous subsection seem
to be sufficiently convincing to justify our method of performing
DWS in nonergodic media, we will now show that the interpretation
of experimental data can be further simplified by introducing
leakage (or absorption) of light at the interface between the
layers in combination with a proper choice of parameters of the
second layer. An optimal choice of the parameters corresponds to
our asymmetric sample B (see Sec.\ \ref{setup}). For this sample,
the optical thickness of the second layer is much smaller than that
of the first one, and the leakage of light at the inter-layer wall
is considerable. This ensures the validity of the multiplication
rule formulated in Sec.\ \ref{mult}, as confirmed by the
experimental results presented below. Figures \ref{multi}(a) and
(b) show the intensity autocorrelation functions $g_2^{(2)}(L_1,
L_2, \tau)-1$ obtained for the two-cell setup (open circles). The
results for two different second cells are displayed. Full squares
show $g_2^{(1)}(L_2, \tau)-1$ obtained for the isolated second
cells, just as in Fig.\ \ref{symmetric}. The corresponding single
scattering correlation time $\tau_0 = (k^2 D_B)^{-1}$ of the
particles in the second cell has been varied by changing the
glycerol content in the suspending liquid, which affects the
viscosity of the liquid and, consequently, the particle diffusion
coefficient $D_B$.

\begin{figure}[t]
\vspace{0.5cm} \psfig{file=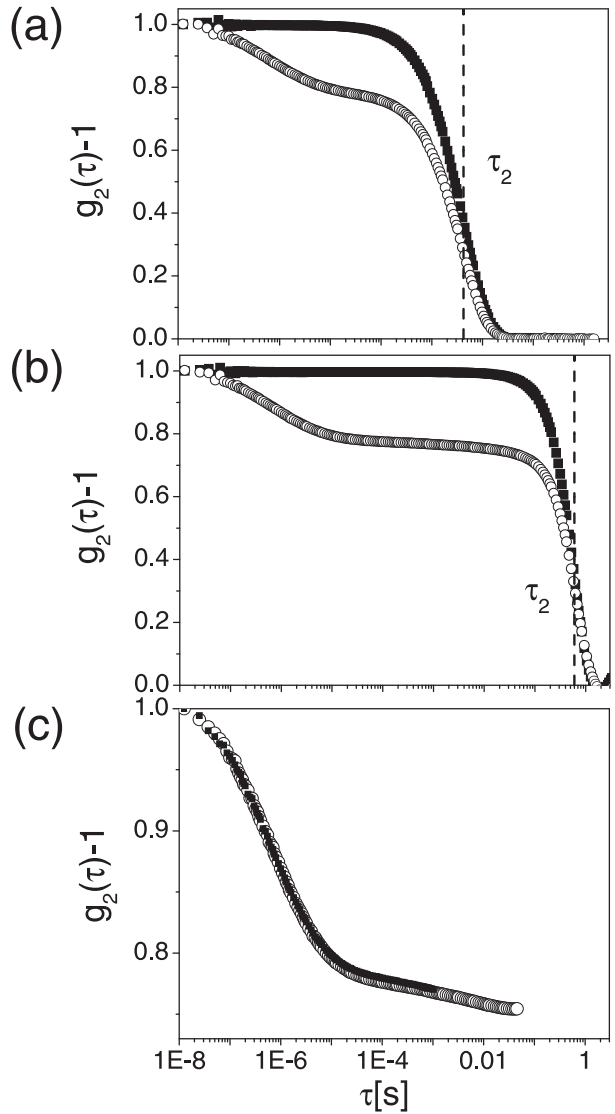,width=8cm} \vspace{0.5cm}
\caption{Multiplication rule:  (a) Intensity autocorrelation
function measured in transmission through the asymmetric two-cell
sample B (open circles). Full squares show the autocorrelation
function measured for the isolated second cell [decay time $\tau_2
= \tau_0 (l_2^*/L_2)^2$]. (b) Same as (a), but for a different
medium inside the second cell (higher glycerol content of the
suspending liquid leading to a reduced particle diffusion
coefficient and a larger value of $\tau_2$). (c) The ratios of the
autocorrelation functions are shown for the data of panels (a)
(open circles) and (b) (full squares). The results are identical
for $\tau < 5 \tau_2$ [with $\tau_2$ of panel (a)], and yield the
ensemble-averaged intensity autocorrelation function $g_2^{(1)}
(L_1,\tau )-1$ corresponding to the nonergodic first cell taken
alone. Inset: Relative difference $\Delta(\tau)= \{ [g_2^{(a)}
(\tau ) - 1] - [g_2^{(b)} (\tau ) - 1] \}/[g_2^{(a)} (\tau ) - 1]$
between the two correlation functions of the main plot.}
\label{multi}
\end{figure}

According to the multiplication rule [Eq.(\ref{multi2})], the ratio
$[g_2^{(2)} (L_1 ,L_2 ,\tau )-1]/[g_2^{(1)} (L_2 ,\tau )-1]$ is
expected to be equal to $g_2^{(1)} (L_1 ,\tau )-1$, i.e. the
autocorrelation function corresponding to the (first) nonergodic
light-scattering cell. To check this prediction, we have calculated
corresponding ratios for the curves of Figs.\ \ref{multi}(a) and
(b). The results are presented in Fig.\ \ref{multi}(c) [full
squares correspond to the data of Fig.\ \ref{multi}(a), while open
circles --- to the data of Fig.\ \ref{multi}(b)]. The two curves
are indistinguishable, supporting the validity of the
multiplication rule. Experimentally, we find that for $\tau < 5
\tau_2$ [with $\tau_2 = \tau_0 (l_2^*/L_2)^2$] the deviations from
the multiplication rule are negligible. The perfect agreement of
the results obtained for two different media in the second cell
confirms the validity of our method.

\begin{figure}[t]
\vspace*{-0.5cm} \psfig{file=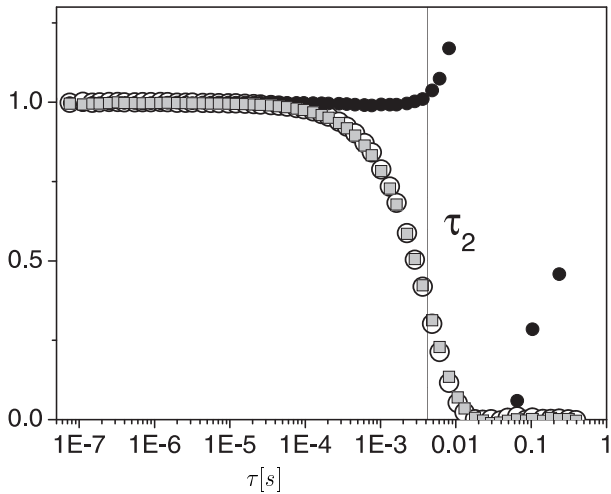,width=8cm} \vspace*{-3.5cm}
\caption{Intensity autocorrelation function for a two-cell sample
with a rigid first layer (the colloidal gel in the first layer was
replaced by a white paper which mimics a rigid turbid medium) (open
circles). Full squares show the autocorrelation function
corresponding to the isolated second layer. The ratio of the two
autocorrelation functions (full circles) is one to a good accuracy
for $\tau < \tau_2 = \tau_0 (l_2^*/L_2)^2$, corresponding to the
absence of scatterer motion in the first layer.} \label{paper}
\end{figure}

As an additional check of the multiplication rule, we have
performed measurements of the intensity autocorrelation function
$g_2^{(2)}(L_1, L_2, \tau)-1$ with the first layer of the
double-layer sample being totally rigid (white paper was used to
model rigid but turbid random medium). The results are presented in
Fig.\ \ref{paper} by open circles for $g_2^{(2)}(L_1, L_2, \tau)-1$
and full squares for $g_2^{(1)}(L_2, \tau)-1$ (the latter is
measured for the isolated second layer). Full circles are obtained
by calculating the ratio $[g_2^{(2)} (L_1 ,L_2 ,\tau
)-1]/[g_2^{(1)} (L_2 ,\tau )-1]$. As expected, the ratio is one
over a sufficiently extended time range (the ratio one corresponds
to no scatterer motion in the paper).

\section{How to apply the two-cell technique ?}
\label{apply} One of the major advantages of the two-cell technique
is its simplicity. Nevertheless, we would like to point out some
pitfalls when designing a two-cell experiment. Firstly, it is
worthwhile to mention the problem of low transmitted light
intensity. Building a sandwich with regular glass cuvettes can
result in a sample of significant overall thickness. Light incident
on the setup spreads out diffusely which leads to relatively low
intensity of light reaching the transmission side. The problem
becomes even more severe if absorption and/or leakage of light are
introduced at the inter-cell wall to simplify the interpretation of
experimental data by using the multiplication rule. Secondly, due
to reflections and additional leakage of diffuse light at the
interface glass walls, some of the light scattered out of the cell
can reach the detector without passing through the whole sample. To
avoid this problem, we recommend to put an aperture (e.g. a black
plastic foil with a centered hole of typically $5$ mm in diameter)
between the cells to suppress photons scattered snake-like along
the cell walls. Ideally, the double-cell sandwich would have only
one moderately absorbing thin interface wall with a built-in
circular aperture forcing all transmitted light to pass through the
whole sample.

\section{Summary and Outlook}
\label{concl} In the present paper we propose and test
experimentally a new method for the application of diffusing-wave
spectroscopy to nonergodic turbid media. We show that light
transmitted trough a sandwich of two turbid samples can be
considered ergodic even if only the second sample is ergodic. The
autocorrelation function of the transmitted intensity can be
quantitatively described by the diffusion theory. This allows
direct application of diffusing-wave spectroscopy for the
characterization of nonergodic media without any additional efforts
usually required to achieve ensemble averaging (i.e. without
translation or rotation of the sample in course of the correlation
function measurement\cite{boas95,skip98,heck98,xue92}). Proper
averaging of the light-scattering signal is ensured by adding a
second, ergodic light-scattering cell with unprecedented accuracy.
In order to simplify the analysis of the experimental data, the
parameters of the double-cell sample can be optimized. Namely,
moderate absorption and/or leakage of light should be introduced at
the interface between the light-scattering cells, and the optical
thickness of the second, ergodic cell should be reduced well below
the optical thickness of the cell containing the nonergodic medium,
while the dynamics of scatterers in the second cell should be
chosen slow. Under these conditions we have shown that the {\em
field\/} autocorrelation function of light transmitted through the
double-layer sample can be written as a product of autocorrelation
functions corresponding to the individual layers: $ g_1^{(2)} (L_1
,L_2 ,\tau ) = g_1^{(1)} (L_1 ,\tau ) \times g_1^{(1)} (L_2 ,\tau
)$. Consequently, it is sufficient to measure the {\em intensity\/}
autocorrelation functions\cite{field} of the two cell setup
$g_2^{(2)} (L_1 ,L_2 ,\tau )$ and of the second, ergodic layer
$g_2^{(1)} (L_2 ,\tau )$, in order to obtain the properly averaged
intensity autocorrelation function of the nonergodic layer
$g_2^{(1)} (L_1 ,\tau )-1 = [g_2^{(2)} (L_1 ,L_2 ,\tau
)-1]/[g_2^{(1)} (L_2 ,\tau )-1]$. Henceforth application of
diffusing-wave spectroscopy for the characterization of nonergodic
media becomes straightforward. Our method is particularly suited
for the study of time evolving systems, e.g. aggregating and
gelling particle suspensions (see also Refs.\ \onlinecite{romer00}
and \onlinecite{wyss01}), since the data acquisition time can
easily be adjusted from a few seconds to several hours. Other
promising applications include DWS-microrheology which has
previously been limited to time and length scales where the motion
of tracer particles is not significantly
constrained\cite{mason95,gisler98}. We expect the two-cell
technique to extend the measurement range of this method and to
provide access to a novel class of (solid-like) materials.

\begin{acknowledgments}
We thank Veronique Trappe for useful comments and discussions.
\end{acknowledgments}


\begin{references}
\bibitem{maret87}
G. Maret and P. E. Wolf, Z. Phys. B {\bf 65,} 409 (1987).

\bibitem{pine88}
D. J. Pine, D. A. Weitz, P. M. Chaikin, and E. Herbolzheimer, Phys.
Rev. Lett. {\bf 60,} 1134 (1988).

\bibitem{pine89} D.  J.  Pine, D.  A.  Weitz, G.  Maret, P.
E.  Wolf, E Herbolzheimer, and P.  M.  Chaikin, in:  {\em
Scattering and Localization of Classical Waves in Random Media,}
ed.  P.  Sheng (World Scientific, Singapore, 1989), p.  312.

\bibitem{pine90a} D.  J.  Pine, D.  A.  Weitz, J.  X.  Zhu,
and E.  Herbolzheimer, J.  Phys.  France {\bf 51,} 2101 (1990).

\bibitem{weitz93a} D.  A.  Weitz and D.  J.  Pine, in Ref.\
\onlinecite{brown93}, p.  652.

\bibitem{weitz93b}
D. A. Weitz, J. X. Zhu, D. J. Durian, H. Gang, and D. J. Pine,
Phys. Scr. {\bf T49B,} 610 (1993).

\bibitem{kaplan94} P. D. Kaplan, A. D. Dinsmore, A. G. Yodh, and D. J. Pine,
Phys. Rev. E. {\bf 50,} 4827 (1994).

\bibitem{berne76} B.  J.  Berne and R.  Pecora, {\em Dynamic
Light Scattering\/} (Wiley, New York, 1976).

\bibitem{brown93} {\em Dynamic Light Scattering,} ed.  W. Brown
(Oxford University Press, New York, 1993).

\bibitem{maret97}
G. Maret, Curr. Opin. Coll. Int. Sci. {\bf 2,} 251 (1997).

\bibitem{romer00}
S.  Romer, F.  Scheffold, and P. Schurtenberger, Phys. Rev. Lett. {\bf 85,}
4980 (2000).

\bibitem{yodh93} P. D. Kaplan, A. G. Yodh and D. F. Townsend, J. Coll. Int. Sci {\bf
155}, 319 (1993).

\bibitem{urbanb00}
C. Urban, S. Romer, F. Scheffold, and P. Schurtenberger, Progr.
Colloid Polym. Sci. {\bf 115,} 270 (2000).

\bibitem{urban00}
C. Urban, S. Romer, and P. Schurtenberger, in {\em Proceedings of
the 2nd International Symposium on Food Rheology and Structure,} P.
Fischer, I. Marti, and E. J. Windhab, Eds., (Zurich, 2000), p. 41.

\bibitem{wyss01} H. Wyss, S.Romer, F. Scheffold, P. Schurtenberger,
and L. J. Gauckler, submitted.

\bibitem{gang95}
H. Gang, A. H. Krall, and D. A. Weitz, Phys. Rev. E {\bf 52,} 6289
(1995).

\bibitem{hebraud97}
P. Hebraud, F. Lequeux, J. P. Munch, and D. J. Pine, Phys. Rev.
Lett. {\bf 78,} 4657 (1997).

\bibitem{lisy99} V. Lisy and B. Brutovsky, Phys. Rev. E {\bf 59,} 3765 (1999)
(comment on Ref.\ \onlinecite{gang95}).

\bibitem{durian91}
D. J. Durian and D. A. Weitz, Science {\bf 252,} 686 (1991); D. J.
Durian, D. A. Weitz, and D. J. Pine, Phys. Rev. A {\bf 44,} R7902
(1991).

\bibitem{earnshaw94}
J. C. Earnshaw and A. H. Jaafar, Phys. Rev. E {\bf 49,} 5408
(1994).

\bibitem{gopal95}
A. D. Gopal and D. J. Durian, Phys. Rev. Lett. {\bf 75,} 2610
(1995).

\bibitem{hohler97}
R. H\"ohler, S. Cohen-Addad, and H. Hoballah, Phys. Rev. Lett. {\bf
79,} 1154 (1997).

\bibitem{gopal97}
A. D. Gopal and D. J. Durian, J. Opt. Soc. Am. A {\bf 14,} 150
(1997).

\bibitem{menon97}
N. Menon and  D. J. Durian, Phys. Rev. Lett. {\bf 79,} 3407 (1997);
Science {\bf 275,} 1920 (1997).

\bibitem{durian00}
D. J. Durian, J. Phys. C: Cond. Matt. {\bf 12,} A507 (2000).

\bibitem{boas97a}
D. A. Boas, G. Nishimura, and A. G. Yodh, Proc. SPIE {\bf 2979,}
468 (1997).

\bibitem{palmer99}
A. Palmer, T. G. Mason, X. Jingyuan, S. C. Kuo, and D. Wirtz,
Biophys. J. {\bf 76,} 1063 (1999).

\bibitem{lohwasser99}
R. Lohwasser and G. Soelkner, Appl. Opt. {\bf 38,} 2128 (1999).

\bibitem{boas95} D.  A.  Boas, L.  E.  Campbell, and A.  G.
Yodh, Phys.  Rev.  Lett.  {\bf 75,} 1855 (1995).

\bibitem{heck96} M.  Heckmeier and G.  Maret,
Europhys. Lett. {\bf 34,} 257 (1996).

\bibitem{skip96} S.  E.  Skipetrov and R.  Maynard, Phys.
Lett.  A {\bf 217,} 181 (1996).

\bibitem{heck97} M.  Heckmeier, S.  E.  Skipetrov, G.
Maret, and R.  Maynard, J.  Opt.  Soc.  Am.  A {\bf 14,} 185
(1997).

\bibitem{boas97} D.  A.  Boas and A. G.  Yodh, J.  Opt.  Soc.
Am.  A {\bf 14,} 192  (1997).

\bibitem{skip98} S.  E.  Skipetrov and I.  V.  Meglinskii,
Zh.  \'Eksp.  Teor.  Fiz. {\bf 113,} 1213 (1998) [JETP {\bf 86,}
661 (1998)].

\bibitem{heck98} M.  Heckmeier and G.  Maret, Progr. Colloid
Polym. Sci. {\bf 104,} 12 (1997); Opt. Commun. {\bf 148,} 1 (1998).

\bibitem{mason95} T. G. Mason and D. A. Weitz, Phys. Rev. Lett. {\bf 74}, 1250
(1995).

\bibitem{gisler98} T. Gisler and D. Weitz, Curr. Opin. Coll. Int. Sci. {\bf 3,} 586
(1998).

\bibitem{crocker00} J. C. Crocker, M. T. Valentine, E. R. Weeks, T. Gisler,
P. D. Kaplan, A. G. Yodh, and D. A. Weitz, Phys. Rev. Lett. {\bf
85,} 888 (2000).

\bibitem{mason00} T. G. Mason, T. Gisler, K. Kroy, E. Frey, and D. A. Weitz,
J. Rheol., to appear (2000).

\bibitem{weitz89}D. A. Weitz, D. J. Pine, P. N. Pusey, and R. J. A. Tough,
Phys. Rev. Lett. {\bf 63,} 1747 (1989).

\bibitem{pine90b} D.  J.  Pine, X.  L.  Wu, P.  M.  Chaikin,
and D.  A.  Weitz, J.  Opt.  Soc.  Am.  B {\bf 7,} 15 (1990).

\bibitem{bicout91} D.  Bicout, E.  Akkermans, and R.
Maynard, J.  Phys.  I (Paris) {\bf 1,} 471 (1991).

\bibitem{bicout93} D.  Bicout and R.  Maynard, Physica A
{\bf 199,} 387 (1993).

\bibitem{bicout95} D. J. Bicout and R.  Maynard,
Physica B {\bf 204,} 20 (1995).

\bibitem{skip00}
S. E. Skipetrov and R. Maynard, Phys. Rev. Lett. {\bf 85,} 736
(2000).

\bibitem{skip01}
S. E. Skipetrov, Phys. Rev. E., to appear (2001); see also cond-mat/0101059.

\bibitem{pusey89} P.  N.  Pusey and W.  van Megen, Physica A
{\bf 157,} 705 (1989).

\bibitem{joosten90}
J. G. H. Joosten, E. T. F. Gelad\'e, and P. N. Pusey, Phys. Rev. A
{\bf 42,} 2161 (1990).

\bibitem{megen91}
W. van Megen, S. M. Underwood, and P. N. Pusey, Phys. Rev. Lett.
{\bf 67,} 1586 (1991).

\bibitem{xue92} J.  Z.  Xue, D.  J.  Pine, S.  T.  Milner,
X.-l.  Wu, and P.  M.  Chaikin, Phys.  Rev.  A {\bf 46,} 6550
(1992).

\bibitem{schatzel93}
K. Schatzel, Appl. Opt. {\bf 32,} 3880 (1993).

\bibitem{kirsch96}
S. Kirsch, V. Frenz, W. Schartl, E. Bartsch, and H. Sillescu, J.
Chem. Phys. {\bf 104}, 1758 (1996).

\bibitem{bartsch97}
E. Bartsch, V. Frenz, J. Baschnagel, W. Schartl, and H. Sillescu,
J. Chem. Phys. {\bf 106,} 3743 (1997).

\bibitem{luca2000}
L. Cipelletti, S. Manley, R.C. Ball, and D. A. Weitz, Phys. Rev.
Lett. {\bf 84,} 2275 (2000).

\bibitem{patent2000} A patent application for this method
was filed with the Suisse Institute of Intellectual Property on
February 27, 2000 under number 200 0335/00.

\bibitem{field}
The field and intensity autocorrelation functions of
multiple-scattered waves are directly related through the Siegert
relation, see Ref.\ \onlinecite{berne76} and Eq.\ (\ref{siegert}).

\bibitem{gang99} H. Gang, A. H. Krall, H. Z. Cummins and D. A. Weitz,
Phys. Rev. E. {\bf 59}, 715 (1999).

\bibitem{nisato00}
G. Nisato, P. Hebraud, J.-P. Munch, and S. J. Candau, Phys. Rev. E
{\bf 61,} 2879 (2000).

\bibitem{harden2000}
A. Knaebel, M. Bellour, J.-P. Munch, V. Viasnoff, F. Lequeux, and
J. L. Harden, Europhys. Lett. {\bf 52,} pp. 73-79 (2000)

\bibitem{zhu91}
J. X. Zhu, D. J. Pine, and D. A. Weitz, Phys. Rev. A {\bf 44,} 3948
(1991).

\bibitem{haskell94}
R. C. Haskell, L. V. Swaasand, T. Tsay, T. Feng, M. S. McAdams, and
B. J. Tromberg, J. Opt. Soc. Am. A {\bf 11,} 2727 (1994).

\bibitem{aronson95}
R. Aronson, J. Opt. Soc. Am. A {\bf 12,} 2532 (1995).

\bibitem{chen98}
N. G. Chen and J. Bai, Phys. Rev. Lett. {\bf 80,} 5321 (1998).

\bibitem{vera96}
M. U. Vera and D. J. Durian, Phys. Rev. E {\bf 53,} 3215 (1996).

\bibitem{correlation} We note that in the case of relatively small particles
($k_{0} d \approx 1$) structural correlations $S(q)$ and
hydrodynamic interactions $h(q)$ have to be taken into account if
present. This complicates a quantitative determination of the
particle mean square displacement significantly. For details see
Ref.\ \onlinecite{weitz93a}.

\bibitem{krall98} A.  H.  Krall and D.  A.  Weitz, Phys.
Rev.  Lett.  {\bf 80,} 778 (1998).

\bibitem{gauckler94} T.  J.  Graule, F.  H.  Baader, and L.
J.  Gauckler, J.  Mater.  Educ.  {\bf 16,} 243 (1994)

\bibitem{Fraden90} S.  Fraden and G.  Maret, Phys.  Rev.
Lett.  {\bf 65,} 512 (1990); X.  Qiu, X.  L.  Wu, J.  Z. Xue, D. J.
Pine, D.  A. Weitz and P.  M. Chaikin, Phys. Rev.  Lett. {\bf 65},
516 (1990).

\bibitem{shapiro86} B.  Shapiro, Phys.  Rev.  Lett. {\bf 57,} 2168 (1986).

\bibitem{ripoll00}
J. Ripoll, M. Nieto-Vesperinas, S. R. Arridge, and H. Dehghani, J.
Opt. Soc. Am. A {\bf 17,} 1671 (2000).

\bibitem{bound}
See Eq.\ (36) of Ref.\ \onlinecite{ripoll00}, where we replace
$U_1(\vec{r})$ by $G_1(\vec{r}, \tau) \equiv G_1(z, \tau)$ and
introduce $f(\vec{r}) = (1/\pi) \int_S {\mathcal G}(\vec{r} -
\vec{r}^{\prime}) {\rm d} S^{\prime}$, where the integration is
over one of the surfaces ($z = L_1$ or $z = L_1 + \Delta$) of the
nonscattering wall separating the turbid layers, and ${\mathcal
G}(\vec{r} - \vec{r}^{\prime})$ is defined by Eq.\ (18) of Ref.\
\onlinecite{ripoll00}.

\bibitem{festimate}
We neglect the absorption of light in the glass wall, and assume
that $f$ is mainly determined by the leakage of light in transverse
directions [see Eq.\ (\ref{fleak})].

\end{references}
\end{document}